# Uncovering the lowest thickness limit for room-temperature ferromagnetism of Cr$_{1.6}$Te$_2$

Sandeep Kumar Chaluvadi,[*,†] Shyni Punathum Chalil,[†,‡] Anupam Jana,[†,‡] Deepak Dagur,[†,¶] Giovanni Vinai,[†] Federico Motti,[†] Jun Fujii,[†] Moussa Mezhoud,[§] Ulrike Lüders,[§] Vincent Polewczyk,[†,∥] Ivana Vobornik,[†] Giorgio Rossi,[†,⊥] Chiara Bigi,[#] Younghun Hwang,[*,@] Thomas Olsen,[*,△] Pasquale Orgiani,[*,†] and Federico Mazzola[*,†,∇]

†CNR-IOM Istituto Officina dei Materiali, Area Science Park, I-34149 Trieste, Italy

‡International Centre for Theoretical Physics (ICTP), Str. Costiera 11, I-34151 Trieste, Italy

¶Department of Physics, University of Trieste, Trieste, Via Alfonso Valerio 2, 34127, Italy

§Normandie Univ ENSICAEN UNICAEN CNRS CRISMAT 14000 Caen, France

∥Groupe d'Etude de la Matière Condensée (GEMaC), Université Paris-Saclay, UMR 8635 Université deVersailles Saint-Quentin en Yvelines & CNRS, Versailles, France

⊥Dipartimento di Fisica, Università degli studi di Milano, IT-20133 Milano, Italy

#Synchrotron SOLEIL, F-91190 Saint-Aubin, France

@Electricity and Electronics and Semiconductor Applications, Ulsan College, Ulsan 44610, Republic of Korea

△CAMD, Computational Atomic-Scale Materials Design, Department of Physics, Technical University of Denmark, 2800, Kongens Lyngby, Denmark

∇Department of Molecular Sciences and Nanosystems, Ca' Foscari University of Venice, I-30172 Venice, Italy

E-mail: chaluvadi@iom.cnr.it; younghh@uc.ac.kr; tolsen@fysik.dtu.dk; pasquale.orgiani@cnr.it; federico.mazzola@unive.it

**Abstract**

Metallic ferromagnetic transition metal dichalcogenides have emerged as important building blocks for scalable magnonics and memory applications. Downscaling such systems to the ultra-thin limit is critical to integrate them into technology. Here, we achieved layer-by-layer control over the transition metal dichalcogenide $Cr_{1.6}Te_2$ by using pulsed laser deposition, and we uncovered the minimum critical thickness above which room temperature magnetic order is maintained. The electronic and magnetic structure is explored experimentally and theoretically and it is shown that the films exhibit strong in-plane magnetic anisotropy as a consequence of large spin-orbit effects. Our study elucidates both magnetic and electronic properties of $Cr_{1.6}Te_2$, and corroborates the importance of intercalation to tune the magnetic properties of nanoscale materials architectures.

The discovery of two-dimensional layered ferromagnets has given researchers the opportunity to study new physical magnetic phenomena as well as provided the community with a new toolbox for spintronics.[1–6] However, applications remain limited, because the vast majority of layered materials exhibits ferromagnetism only at cryogenic temperatures.[7,8] Investigating novel ferromagnetic materials, in terms of both lower thickness limit and transition temperature is then crucial. The transition metal dichalcogenide $CrTe_2$ has often been proposed as a candidate for technological applications, because it can be exfoliated easily, and even in 10 nm flakes ferromagnetic behaviour above room temperature has been recorded.[9–11] Recently, Cr-intercalation between consecutive $CrTe_2$ layers has been shown to be a way to chemically tune the magnetism, electronic properties, and topology.[12–14] In previous studies of Cr-intercalated transition metal dichalcogenides, a vast array of magnetic arrangements depending on the amount of intercalated Cr was discovered.[13,15] Thus, studies on ferromagnetic systems are desired to engineer efficient high-temperature spin-orbit torque spintronics, such as magneto-resistive random-access memories.[16] Here we focus on $Cr_{1.6}Te_2$, a member of the self-intercalated $CrTe_2$-based compounds.[12]

The phase diagram of the $Cr_{1+\delta}Te_2$ class of materials is rich.[12] Across the series, $Cr_{1.6}Te_2$ is one of the most structurally stable phases, exhibiting room temperature ferromagnetism in the bulk.[17] Here, we demonstrate the ability to control the thickness of $Cr_{1.6}Te_2$ in a layer-by-layer fashion. Then, we investigate the electronic and magnetic properties of this system, and ferromagnetism is shown to persist down to a thickness of 5 nm. By means of angle-resolved photoelectron spectroscopy and density functional theory (DFT) calculations, we identify the majority and minority bands conferring the ferromagnetic order to the sample, and we reveal a significant magnetic anisotropy induced by strong SOC effects.

Epitaxial $Cr_{1.6}Te_2$ thin films were grown by pulsed laser deposition (PLD) utilizing a Nd:YAG laser source operating at its first harmonics ($\lambda$=1064 nm) falling in the infrared regime.[18,19] Such a laser ablates a rotating stoichiometric polycrystalline $Cr_{1.6}Te_2$ target (purity 99.99%) placed at a distance of 9.5 cm from the substrate (Fig.1a). This mechanism allows for a few nanometers-thick materials to be synthesized with the precision of a single mono-layer. $Al_2O_3$ (0001)-oriented substrates were chosen because of their dielectric properties, useful for electronics. They were kept at a temperature of 840 K and at the base pressure of $5 \times 10^{-8}$ mbar throughout the growth. This process allows us to build architectures with various thicknesses, from bulk-like to thin-film samples (See schematic in Fig.1a).

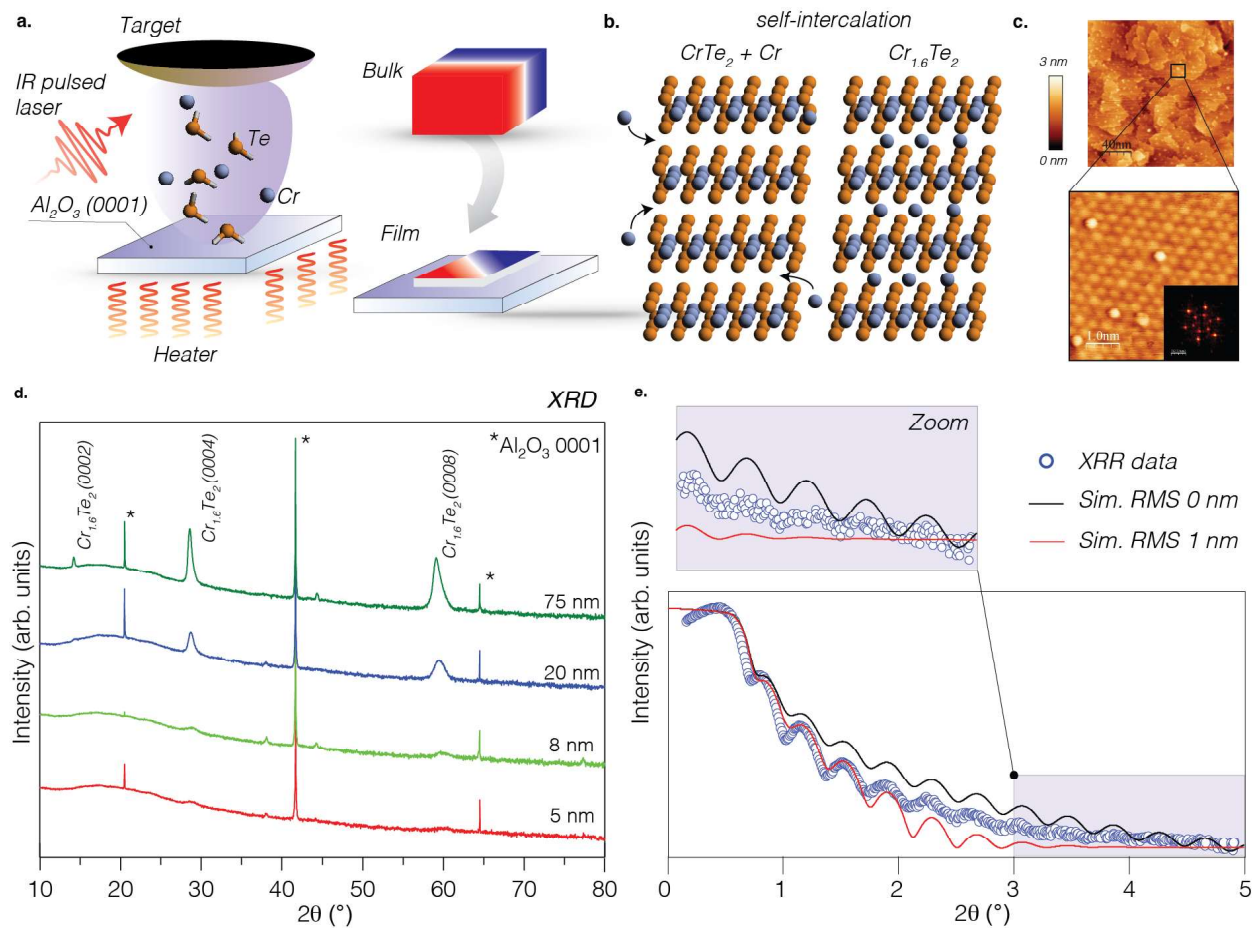

Figure 1: **Growth of self-intercalated $Cr_{1.6}Te_2$.** **a** Pulsed laser deposition: the target material is ablated by an infrared pulsed laser and the constituents deposited on a hot substrate. This technique allows us to control the size of our materials with layer-by-layer precision. In this case, a bulk magnet (red and blue for the magnetic poles) is grown into a thin film form. **b** Cartoon of self-intercalation: the parent compound $CrTe_2$ is a layered van der Waals system with a significant inter-layer spacing. **c** STM topography image of the $Cr_{1.6}Te_2$ film with the zoomed portion in the scan area of $5 \times 5\,nm^2$ showing well-ordered atoms with defect-free sites (in the inset, the corresponding FFT shows a hexagonal symmetry). **d** X-ray diffraction measurements confirming the growth of $Cr_{1.6}Te_2$ in single crystal form on the top of the insulating compound $Al_2O_3$. **e** X-ray reflectivity measurements showing the typical oscillatory behaviour (simulations of XRR curves with surface roughness values of $0\,nm$ (black) and $1\,nm$ (red) are also reported as solid lines).

To characterize the crystal structure and the lattice parameters of the films, thickness-dependent X-ray diffraction (XRD) was performed (See methods). Clear $Cr_{1.6}Te_2$ (000$l$) peaks were observed down to the ultra-thin film limit (see Fig.1d) with intensities decreasing as the thickness gets lower. XRD spectra show no sign of spurious phases or different $Cr_{1.6}Te_2$ crystallographic orientations but solely features belonging to the substrate and to (000$l$)-oriented $Cr_{1.6}Te_2$.[17,20] This is consistent with high crystal quality and with a single-phase. The out-of-plane c-axis lattice parameter calculated from the 0004-diffraction peak is 12.496 ± 0.003Å, as also reported in literature[17,20,21] (see supplementary details phase identification section) thus confirming the phase of films to be $Cr_{1.6}Te_2$. X-rays were also used to estimate the thickness and the surface roughness of the films: for example, for the 20 nm thick crystal, the low angle X-ray reflectivity (XRR) curve shows a very well-defined oscillatory behaviour for $2\theta$ values up tp $4^o$, while, above this angle, the oscillations fall below the experimental sensitivity of the X-ray diffractometer (See in Fig.1e). The surface root-mean-square (RMS) roughness of the films was estimated to be smaller than 1 nm, corresponding to less than one single $Cr_{1.6}Te_2$ unit cell[22] (See Methods). The actual surface roughness of the grown films was additionally probed with precision by *in-situ* scanning tunneling microscopy (STM), yielding an RMS value of 0.35 nm (i.e. well below the thickness of a single unit cell of $Cr_{1.6}Te_2$, Fig.1c), in agreement with XRR.

We now focus on the magnetism of $Cr_{1.6}Te_2$. The MOKE signal from $Cr_{1.6}Te_2$ thin films was collected at 300 K along the [10$\bar{1}$0] crystallographic direction (See Fig.2a). The experiment was performed in longitudinal mode, i.e., the external applied magnetic field is in the plane of the sample and parallel to the plane of the incident light, as shown in the cartoon picture of Fig.2b. A clear hysteresis, typical of ferromagnetic behaviour, is observed down to 8 nm at room temperature, whereas an almost paramagnetic response is found at 5 nm (green curve, Fig.2a - note the extremes of the green curve have been removed to better visualize the hysteresis). Still, at 5 nm small fraction of the initial ferromagnetic order is observed. The coercive field ranges from 14 Oe to 1 Oe, from the thickest to the thinnest compounds. This is a relatively small value compared to most known ferromagnets, and, certainly, a desired property for controlling the magnetic response in spintronic applications.

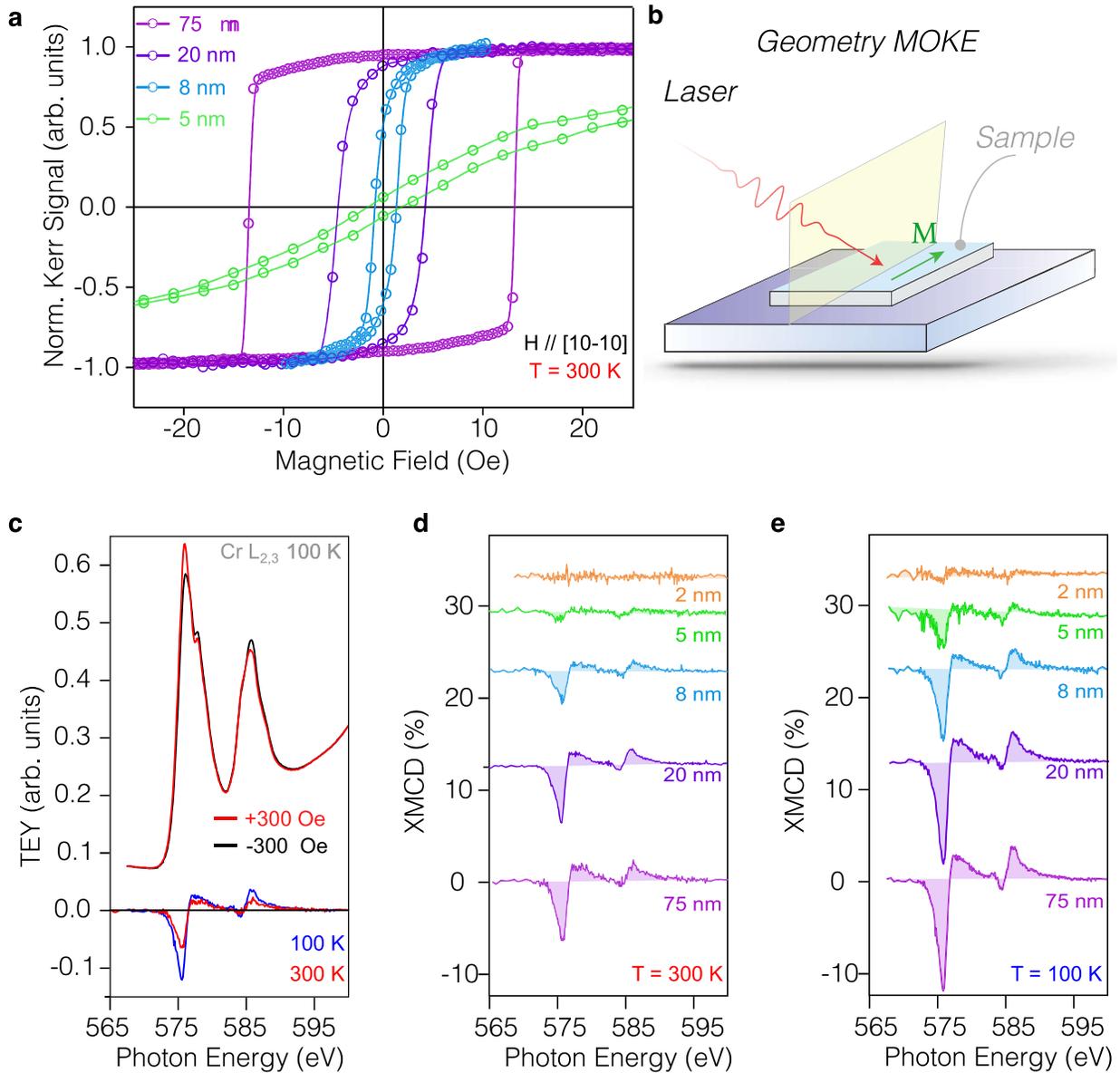

Figure 2: **Cr$_{1.6}$Te$_2$: ultra-thin ferromagnet.** **a** MOKE hysteresis loops collected in longitudinal mode at 300 K along [10$\bar{1}$0] crystallographic direction. Several thicknesses of Cr$_{1.6}$Te$_2$ are shown down to 5 nm. **b** schematic of the longitudinal MOKE setup. **c** Cr L$_{2,3}$ XAS of a 75 nm thick Cr$_{1.6}$Te$_2$ taken at 100 K and the corresponding XMCD signal at 100 and 300 K respectively. **d** XMCD thickness-dependence collected across the Cr L$_{2,3}$ edges both at 300 K and **e** 100 K.

While the minimum thickness limit for room temperature ferromagnetism is found between 5 nm and 8 nm, for smaller thicknesses, the onset of the ferromagnetic ordering occurs at lower temperatures. For films of 2 nm, for instance, at 100 K a clear MOKE hysteresis loop is seen (See Supplementary information figure S10). In addition, for all the films investigated, isotropic in-plane magnetic behavior is observed (see supplementary figures S8, S9, S10). We stress that this particular composition ($\delta$=0.6) is well-known to have a pure in-plane magnetization with no out-of-plane contribution.[12] Importantly, we also performed SQUID measurements (See supplementary figure S8) to access the transition temperature of the system, and apart from a typical ferromagnetic hysteris we found that the magnetization curve has a rapid drop at around 350 K, consistent with the aforementioned room temperature magnetism.

In order to investigate the local electronic character and magnetic ground state of the $Cr_{1.6}Te_2$ films, X-ray absorption spectroscopy (XAS) and X-ray magnetic circular dichroism (XMCD) measurements at the Cr $L_{2,3}$ absorption edges were performed (see methods section for details). The Cr $L_2$ and $L_3$ edges emerge at 585.7 eV and 575.9 eV, respectively (Fig.2c), with spectral features in accordance with the expected $Cr^{2+}/Cr^{3+}$ ratio.[23] The same XAS features are reported for all thicknesses. This indicates a well defined electronic character from the thinnest to the thickest crystal (see supplementary figure S4).

By comparing XMCD spectra at 300 K (Fig.2d) with those at 100 K (Fig.2e), the temperature evolution of the magnetism is revealed: a finite magnetic dichroism is measured across the Cr $L_{2,3}$ edges down to 8 nm at room temperature, consistent with the MOKE results. This is also seen in the 5 nm film at 100 K. Both XAS and XMCD indicate that Cr is responsible for the ferromagnetic order and very little hybridization with Te is found. Importantly, in self-intercalated compounds, the interstitial Cr atoms may be either aligned parallel or antiparallel to the main magnetization of the $CrTe_2$ layers.

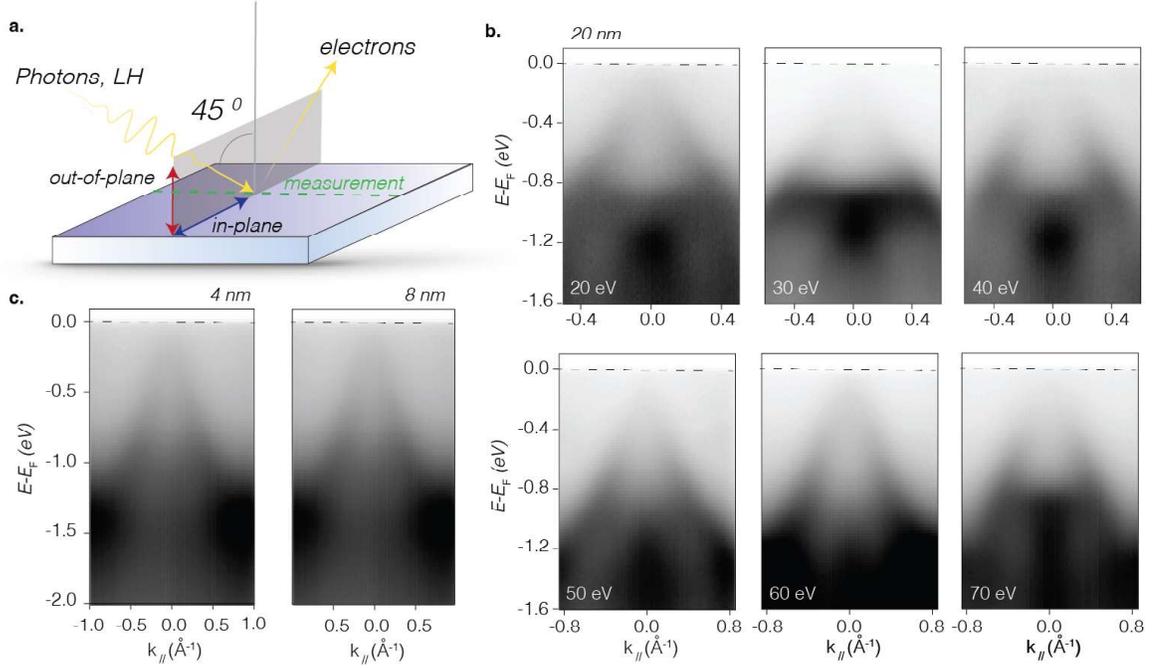

Figure 3: **Electronic structure of $Cr_{1.6}Te_2$.** **a** Photoelectron spectroscopy methodology and experimental geometry: The photons are shone on the sample with a 45° incidence angle. The light is linearly polarized with the polarization plane being orthogonal to the surface. **b** The photoemission shown are collected for 20 nm thickness $Cr_{1.6}Te_2$, and for various photon energies. These films are found to have the highest signal-to-noise ratio among all thicknesses studied. The electronic structure is overall similar at all photon energies, with the exception of changes at approximately 1 eV. This might be either due to bulk bands or redistribution of the spectral weight as an effect of the photoemission matrix elements. **c** photoemission measurements performed on 4 nm and 8 nm thin films $Cr_{1.6}Te_2$ (59 eV and linear horizontal light polarization). The spectra are similar and no noticeable differences are observed. They are characterized by a narrow valence band touching the Fermi surface. From the spectra collected for other thicknesses, we never see sizeable changes in the electronic structure.

Combining photoemission with DFT calculations, we now identify majority and minority characters of the bands and the main orbital contributions to the magnetic order. Photoemission data were collected by using a DA30 hemispherical analyzer at the APE-LE laboratory[24] (Elettra synchrotron - See methods). The experimental electronic structure consists primarily of hole-like bands dispersing up to the Fermi level (Fig.3b-c). The broadening of the electronic states observed is ubiquitous for all thicknesses. In magnetic compounds even when of quasi van der Waals nature, the interlayer coupling is generally stronger than in non-magnetic layered systems and the large orbital overlap is cause of extra broadening. Additional reasons for the band broadening might be attributed to many-body effects, such as electron-electron and electron-impurity scattering, which can contribute to smear out the electronic dispersion.[25–27] Importantly, even the intercalated Cr atoms might present a degree of disorder, and this contributes to broadening too.

The quasi van der Waals nature of the films is nevertheless highlighted by the absence of changes in the spectra collected at various thicknesses, as for example from 20 nm to 4 nm, as shown in Fig.3b-c (if the interactions changed with thickness, different orbital hybridization would be expected). The weak interaction between layers is also unveiled by the lack of a $k_z$ dispersion: by varying photon energy (See Fig.3b and also in supplementary figure S15) the electronic structure remains almost the same and the only visible effect is a redistribution of the spectral weight and a small 'wiggle'. This is also visible in supplementary figure S15, where the momentum distribution curves collected at several values of photon energy do not show significant dispersion and the peak position does not vary across the range of the energies used significantly. Generally, this behaviour is typical of quasi two-dimensional systems.

If the electronic structure does not disperse along $k_z$, a richer scenario occurs in the parallel component of momentum: dispersing spectral weight is found at momenta of approximately 0.75 Å$^{-1}$ just below the Fermi energy (See Fig.4a). These features were not reported previously for bulk compounds, likely due to their extremely weak spectral intensity. To help their visualization, we performed the curvature plots of the photoemission spectra (See Ref.,[28] Fig.4b).

The photoemission experiment is consistent with the DFT calculations of the band structure, which included the ferromagnetic order as well as spin-orbit coupling (Fig. 4c-d). The calculations reveals the valence band of $Cr_{1.6}Te_2$ dispersing up to the Fermi level. The majority of the spectral weight is concentrated around the zone edges M (and K), and the most prominent feature of the valence band has its maximum at the centre of the Brillouin zone and exactly at the Fermi level. This is also visible from the photoemission data curvature plot of Fig.4b. Similarly to the experiment, also the DFT band structure shows weak spectral features at approximately -0.5 eV along the Γ-M path, as indicated by the red box in Fig.4c. Similarly to the photoemission results, such features appear to be the weakest in intensity.

Importantly, there is a technical intrinsic challenge when dealing with self-intercalated compounds. Indeed, for the sake of the calculations, the interstitial Cr atoms need to be ordered within the single unit cell (See Supplementary information for crystal structure and unit cell used in DFT). In a real thin film growth, the intercalated Cr will necessarily have a certain degree of randomness concerning its position within the $CrTe_2$ layers. This leads to an increased broadening in the experimental bands, especially those with a prominent Cr character. In this regard, elucidating the orbital character is crucial: within the energy range considered for the photoemission spectra, the bands exhibit a mixed Cr-Te character, mainly Cr-derived over the full range and mixed Cr-Te character around the Fermi level (See Fig.4e). Cr is the main driver of ferromagnetism, with the majority spin within the whole experimental range shown. In particular, the Cr belonging to the CrTe$_2$ planes contributes with magnetic moments of 3.2-3.3 $\mu_B$ while the intercalated ones give somewhat larger, 3.4 $\mu_B$ (See details of the ferromagnetic properties obtained from DFT in the methods). This is also in agreement with the XMCD data, where no significant magnetic hybridization could be detected. Note that from DFT, the most favorable energetic configuration is obtained when all of the magnetic moments are aligned along the in-plane direction. Even when canting is included or a small degree of out-of-plane plane component is considered, the energy increases significantly, ruling out other possible magnetic configurations for both the Cr atoms of the main CrTe$_2$ planes and the intercalated Cr.

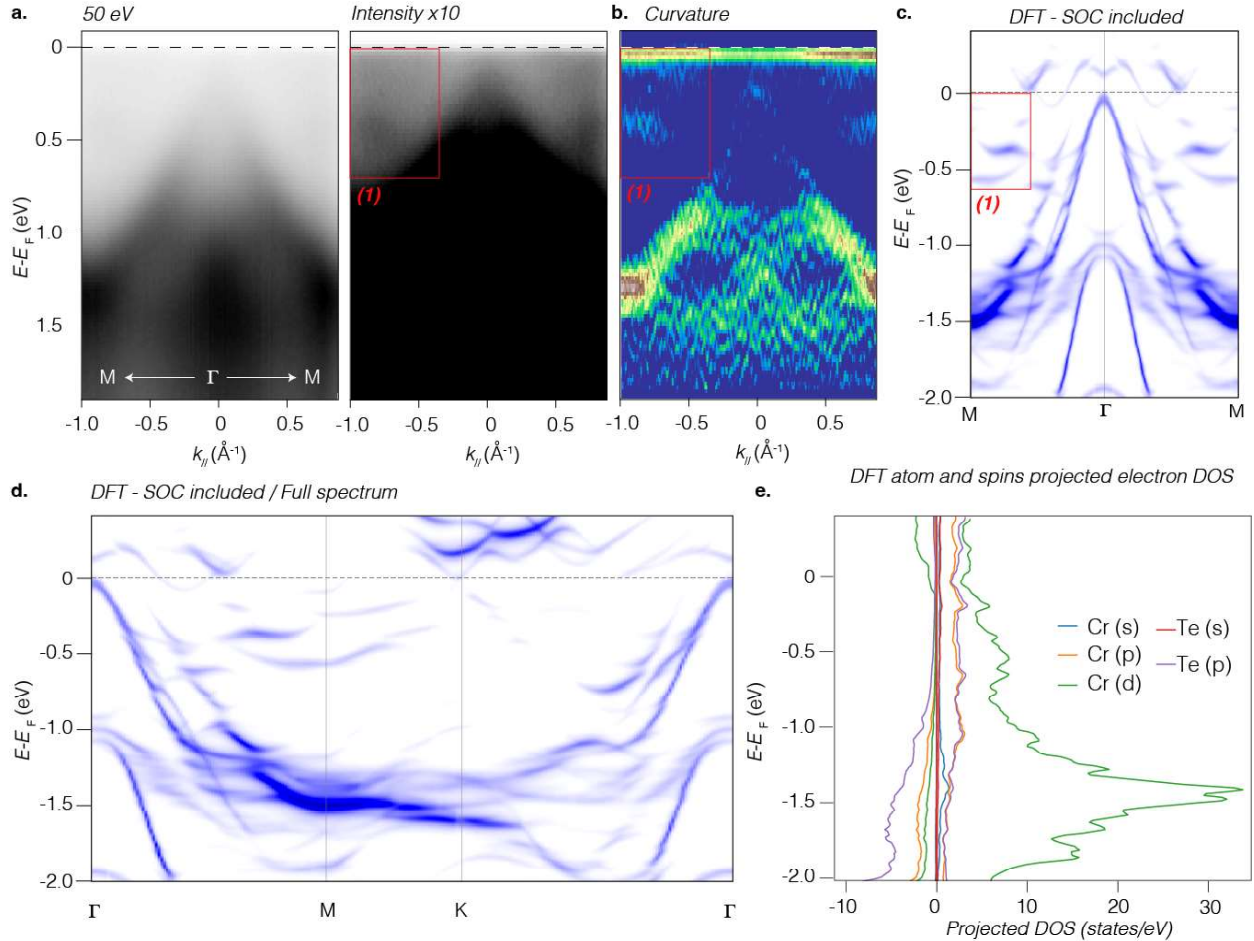

Figure 4: **DFT Calculations for the band structure and comparison with ARPES** **a-c** Photoemission spectra at 50 eV (right, intensity saturated). By saturating the intensity, features matching with a predominantly Cr-character can be tracked. The broadening can be due to the fact that the interstitial Cr has to be assumed ordered in the calculations, however, a certain degree of randomness is present in the experiment. **c** Band structure obtained from DFT along the M-Γ-M direction. The band structures were obtained by unfolding the super cell (required to simulate $Cr_{1.6}Te_2$) to the pristine unit cell of $CrTe_2$. **d** Band structures the Γ-M-K-Γ line. **e** Density of states resolved in different orbital contributions (from DFT). The left hand side (at negative values) shows the minority spin channel with very few states at the Fermi level compared to the majority spin channel (right hand side, positive values).

The Te atoms, despite provide strong spin-orbit coupling to the systems, which is responsible for the large magnetic anisotropy favoring the in-plane ordering, likely to comprise a crucial factor for the large Curie temperature of this compound. According to our calculations, we may refer to the compound as half-metallic in the absence of relativistic effects, since the Fermi surface is mainly comprised by majority carriers. However, when spin-orbit effects are taken into account this picture is no longer rigorous since $S_z$ is no longer a good quantum number and the minority spins may be accessed at the Fermi level, which could have an impact on their applications in spintronics. From the point of view of electronic structure, the spin-orbit coupling generates a drastic band renormalization (See Supplementary Information): not only is the main hole pocket shifted towards the Fermi level of about 500 meV, but also the spectral weight is significantly redistributed as well as the shape of the dispersion is affected. The inclusion of spin-orbit coupling seems to be crucial to describe the experimental results obtained by photoemission. We emphasize that magnetic systems with strong spin orbit coupling might be an important resource to seek for a finite Berry curvature, because the confluence of time-reversal symmetry breaking and spin orbit coupling is at the basis of non-trivial topological properties caused by the Berry curvature itself.[29–31] In this regard, Fujisawa et al.[12] reported that for thicker films of our same composition negative anomalous transport occurs. Thus, even if this goes beyond the scope of our letter, it is interesting to speculate for future works, about investigating thin films materials of $Cr_{1.6}Te_2$ to look for the same phenomenology.

In conclusion, we have unraveled the magnetic and electronic properties in ultrathin films limit of the quasi van der Waals system $Cr_{1.6}Te_2$. The ability of self-intercalating Cr-atoms within the van der Waals gap offers a large degree of tunability of the magnetic order, which survives down to the ultra-thin film limit (5-8 nm). The strong SOC and associated magnetic anisotropy is likely to be a decisive factor for the large Curie temperature, found to be approximately 350 K by SQUID. We finally speculate on possible scenarios for the magnetism below the critical thickness found for room-temperature ferromagnetism (5 nm). Below such a thickness two scenarios may occur: the magnetic moments align out of plane, thus the system will be still ferromagnetic but with a completely different easy axis; the Curie temperature gets drastically reduced and the ferromagnetism cannot be captured within the available temperature range of our instrument. In the light of our work, we are keen to believe that the latter scenario is the most probable, however further investigations which go beyond the scope of this work are needed to establish this. This study encourages future investigations of self-intercalated compounds as a possible viable mean of controlling a large plethora of magnetic orders, down to the nanoscale. Even if temperatures still remain too low for realistic applications, these compounds have shown the ability to modulate their Curie temperature drastically with number of intercalated atoms. Thus, they still offer an important material platform to be considered.

# Methods

**Substrate preparation and thin-film growth optimization:** The c-plane oriented single-crystalline $Al_2O_3$ substrates were commercially purchased from the SurfaceNet Gmbh, Germany. The as-received substrates were subsequently diced into 5×5 mm$^2$ dimensions, rinsed in ethanol and dried under $N_2$ flow, introduced into the PLD uhv chamber with the base pressure of $5\times10^{-8}$ mbar. $Cr_{1.6}Te_2$ thin films were deposited in uhv onto $Al_2O_3$ substrates by using $1^{st}$ harmonic Nd:YAG laser source with a repetition rate of 2 Hz, respectively. The crystalline structure of the deposited films was assessed via XRD, following which, the substrate deposition temperature (Ts) was finely tuned to obtain a flat, defect-free surface. The optimization process was facilitated by *in-situ* STM surface analysis, allowing for precise control over the growth conditions. Films deposited at a temperature of 840 K show flat terraces, low roughness and defect-free surfaces. Whereas, a slight deviation of the Ts leads to an increase in roughness and defect sites on the surface, as discussed in detail in supplementary figure S3.

**Crystal structure:** The XRD measurements were performed in high-resolution mode from Malvern Empyrean high-resolution diffractometer with monochromatized Cu K$\alpha$1 radiation ($\lambda = 1.54056$Å) and a double-bounce Ge(220) crystal in the diffracted beam.

For the surface roughness, the extracted value must be considered as an upper limit for the surface RMS since the fitting procedure is based on a monochromatized X-ray source with negligible lateral inhomogeneities of the beam (while we used a lab-based unmonochromatized X-ray beam).

**STM:** The surface morphology of the films were probed by the *in-situ* STM facility directly connected to the PLD deposition chamber at the APE NFFA facility. The atomic-resolution STM image was acquired with the set point tip current of 2.9 nA and the bias voltage of 240 mV.

**MOKE:** longitudinal Magneto-optic Kerr effect hysteresis loops were measured at the MBE Cluster of the NFFA facility in Trieste[32] using a s-polarized red laser (658 nm) and a photoelastic modulator (PEM) operating at 50 kHz coupled to a lock-in setup, with a laser spot size of about 500x500 $\mu m^2$. Samples were transferred from the PLD deposition chamber to the Cluster via vacuum suitcase so as not to expose the surface of the samples to air and avoid any oxidation/degradation of their properties.[33] The hysteresis loops were measured in UHV conditions, both at 300 and 100 K, cooling the samples with liquid Nitrogen.

**XAS/XMCD:** XAS and XMCD spectra at Cr $L_{2,3}$ edges were taken at the APE-HE beamline of the Elettra synchrotron radiation facility in Trieste.[34] All spectra were measured in total electron yield (TEY) mode by normalizing the intensity of the sample current to the incident photon flux current at each energy value. The sample surface was kept at 45° with respect to the incident beam, for a beam spot of around 200x200 $\mu m^2$. XMCD measurements were performed at magnetic remanence, after applying an in-plane alternating magnetic field pulse of ±300 Oe along the [10$\bar{1}$0] $Al_2O_3$ crystallographic axis at every measured photon energy value. The recorded XMCD signal was then normalized to the sum of the two XAS for positive and negative magnetic field values. All the XMCD signals were shown in %, taking into consideration the angle of 45° between the photon angular momentum and the in-plane magnetization, as well as the 75% circular polarization degree of light. Spectra photon values were aligned by measuring a reference powder during each scan.

**ARPES:** Photoemission was performed at the APE-LE laboratory of the NFFA cluster in Trieste, Elettra synchrotron. The energy and experimental resolutions of the instrument were better than 12 meV and 0.018 Å$^{-1}$, respectively. The samples were transferred under ultrahigh vacuum conditions, without being exposed to air. The temperature was kept at approximately 20 K throughout the data acquisition and the samples aligned along Γ-M. The geometry of the photoemission experiment was as described in Fig.3a, with linear horizontal polarization for the incoming beam probing both in-plane and out-of-plane orbital contributions from the electronic structure. We find this condition to be the optimal one to detect most features of $Cr_{1.6}Te_2$.

**DFT:** The DFT calculations were performed with the electronic structure package GPAW using the projector-augmented wave method and plane waves.[35,36] For the relaxation of the $Cr_{1.6}Te_2$ structure, we used a plane wave cutoff of 800 eV and a $7 \times 7$ uniform $k$-point sampling. A subsequent calculation of frozen phonons in a $2 \times 2$ super cell implies dynamic stability of the structure. The structure was found to reside 28 meV above the convex hull in agreement with the expected high stability of the compound. The calculations yielded a half-metallic state with 13.9 $\mu_B$ per formula unit, with the local moments of the intercalated Cr atoms being slightly larger than the rest.

The structure was modeled by a $\sqrt{5} \times \sqrt{5}$ supercell of $CrTe_2$ where 3 intercalated Cr atoms were added. Then, the structure was fully relaxed resulting in a ferromagnetic metallic ground state. By considering a state with anti-aligned intercalated atoms, a higher energy configuration compared to the fully ferromagnetic state was found (0.3 eV per formula unit), in agreement with the experimental evidence of ferromagnetic alignment of intercalated atoms.

# Acknowledgements

The experiments were performed at the NFFA APE-LE and APE-HE beamlines on the Elettra synchrotron radiation source and supported in the framework of the Nanoscience Foundry and Fine-Analysis (NFFA-MUR Italy Progetti Internazionali) facility. We acknowledge Elettra Sincrotrone Trieste for providing access to APE-HE beamline through Proposal No.20225460. F.M. greatly acknowledges the SoE action of PNRR, number SOE_0000068. This work was supported by the National Research Foundation of Korea (NRF) funded by the Ministry of Education, Science and Technology (NRF-2019M2C8A1057099 and NRF-2022R1I1A1A01063507)

# References

(1) Wang, Q. H. et al. The Magnetic Genome of Two-Dimensional van der Waals Materials. *ACS Nano* **2022**, *16*, 6960–7079.

(2) Gibertini, M.; Koperski, M.; Morpurgo, A. F.; Novoselov, K. S. Magnetic 2D materials and heterostructures. *Nature Nanotechnology* **2019**, *14*, 408–419.

(3) Burch, K. S.; Mandrus, D.; Park, J. G. Magnetism in two-dimensional van der Waals materials. *Nature* **2018**, *563*, 47–52.

(4) Gong, C.; Zhang, X. Two-dimensional magnetic crystals and emergent heterostructure devices. *Science* **2019**, *363, 6428*, eaav4450.

(5) Wang, Z. et al. Electric-field control of magnetism in a few-layered van der Waals ferromagnetic semiconductor. *Nature Nanotechnology* **2018**, *13*, 554–559.

(6) Wang, H.; Li, X.; Wen, Y.; Cheng, R.; Yin, L.; Liu, C.; Li, Z.; He, J. Two-dimensional ferromagnetic materials: From materials to devices. *Applied Physics Letters* **2022**, *121, 220501*.

(7) Gong, C.; Li, L.; Li, Z.; Ji, H.; Stern, A.; Xia, Y.; Cao, T.; Bao, W.; Wang, C.; Wang, Y.; Qiu, Z. Q.; Cava, R. J.; Louie, S. G.; Xia, J.; Zhang, X. Discovery of intrinsic ferromagnetism in two-dimensional van der Waals crystals. *Nature* **2017**, *546*, 265–269.

(8) Huang, B.; Clark, G.; Navarro-Moratalla, E.; Klein, D. R.; Cheng, R.; Seyler, K. L.; Zhong, D.; Schmidgall, E.; McGuire, M. A.; Cobden, D. H.; Yao, W.; Xiao, D.; Jarillo-Herrero, P.; Xu, X. Layer-dependent ferromagnetism in a van der Waals crystal down to the monolayer limit. *Nature* **2017**, *546*, 270–273.

(9) Sun, X. et al. Room temperature ferromagnetism in ultra-thin van der Waals crystals of 1T-CrTe2. *Nano Research* **2020**, *13*, 3358–3363.

(10) Zhang, X. et al. Room-temperature intrinsic ferromagnetism in epitaxial CrTe$_2$ ultrathin films. *Nature Communications* **2021**, *12*, 2492.

(11) Yu, F.; Yin, Y.; Liu, G.; Tian, Q.; Meng, Q.; Zhao, W.; Wang, K.; Wang, C.; Yang, S.; Wu, D.; Wan, X.; Zhang, Y. Thickness-dependent structural phase transition and self-intercalation of two-dimensional ferromagnetic chromium telluride thin films. *Applied Physics Letters* **2022**, *120*, 261602.

(12) Fujisawa, Y.; Pardo-Almanza, M.; Hsu, C.; Mohamed, A.; Yamagami, K.; Krishnadas, A.; Chang, G.; Chuang, F.; Khoo, K. H.; Zang, J.; Soumyanarayanan, A.; Okada, Y. Widely Tunable Berry Curvature in the Magnetic Semimetal Cr$_{1+\delta}$Te$_2$. *Advanced Materials* **2023**, *35*, 2207121.

(13) Fujisawa, Y. et al. Tailoring magnetism in self-intercalated in self-intercalated Cr$_{1+\delta}$Te$_2$ epitaxial films. *Physical Review Materials* **2020**, *4*, 114001.

(14) Chi, H. et al. Strain-tunable Berry curvature in quasi-two-dimensional chromium telluride. *Nature Communications* **2023**, *14, 3222*.

(15) Ou, Y.; Yanez, W.; Xiao, R.; Stanley, M.; Ghosh, S.; Zheng, B.; Jiang, W.; Huang, Y.-S.; Pillsbury, T.; Richardella, A.; Liu, C.; Low, T.; Crespi, V. H.; Mkhoyan, K. A.; Samarth, N. ZrTe2/CrTe2: an epitaxial van der Waals platform for spintronics. *Nature Communications* **2022**, *13*, 2972.

(16) Ishida, K.; Shirokura, T.; Hai, P. N. Enhanced spin Hall effect at high temperature in non-centrosymmetric silicide TaSi2 driven by Berry phase monopoles. *Applied Physics Letters* **2023**, *123*, 262402.

(17) Liu, H.; Fan, J.; Zheng, H.; Wang, J.; Ma, C.; Wang, H.; Zhang, L.; Wang, C.; Zhu, Y.; Yang, H. Magnetic properties and critical behavior of quasi-2D layered Cr$_4$Te$_5$ thin film. *Frontiers of Physics* **2023**, *18*, 13302.

(18) Chaluvadi, S. K.; Mondal, D.; Bigi, C.; Knez, D.; Rajak, P.; Ciancio, R.; Fujii, J.; Panaccione, G.; Vobornik, I.; Rossi, G.; Orgiani, P. Pulsed laser deposition of oxide and metallic thin films by means of Nd:YAG laser source operating at its 1st harmonics: recent approaches and advances. *Journal of Phys. Materials* **2021**, *4*, 032001.

(19) Orgiani, P. et al. Dual pulsed laser deposition system for the growth of complex materials and heterostructures. *Review of Scientific Instruments* **2023**, *94*, 033903.

(20) Wang, W.; Fan, J.; Liu, H.; Zheng, H.; Ma, C.; Zhang, L.; Sun, Y.; Wang, C.; Zhu, Y.; Yang, H. Fabrication and magnetic–electronic properties of van der Waals $Cr_4Te_5$ ferromagnetic films. *CrystEngComm* **2022**, *24*, 674.

(21) Wang, J.; Wang, W.; Fan, J.; Zheng, H.; Liu, H.; Ma, C.; Zhang, L.; Tong, W.; Ling, L.; Zhu, Y.; Yang, H. Epitaxial growth and room-temperature ferromagnetism of quasi-2D layered $Cr_4Te_5$ thin film. *Journal of Physics D: Applied Physics* **2022**, *55*, 165001.

(22) Windt, D. L. IMD—Software for modeling the optical properties of multilayer films. *Computers in Physics* **1998**, *12*, 360.

(23) Ishida, Y.; Kobayashi, M.; Hwang, J.-I.; Takeda, Y.; ichi Fujimori, S.; Okane, T.; Terai, K.; Saitoh, Y.; Muramatsu, Y.; Fujimori, A.; Tanaka, A.; Saito, H.; Ando, K. X-ray Magnetic Circular Dichroism and Photoemission Study of the Diluted Ferromagnetic Semiconductor Zn1-xCrxTe. *Applied Physics Express* **2008**, *1*, 041301.

(24) Bigi, C.; Das, P. K.; Benedetti, D.; Salvador, F.; Krizmancic, D.; Sergo, R.; Martin, A.; Panaccione, G.; Rossi, G.; Fujii, J.; Vobornik, I. Very efficient spin polarization analysis (VESPA): New exchange scattering-based setup for spin-resolved ARPES at APE-NFFA beamline at Elettra. *Journal of Synchrotron Radiation* **2017**, *24*, 750–756.

(25) Mazzola, F.; Wells, J. W.; Yakimova, R.; Ulstrup, S.; Miwa, J. A.; Balog, R.; Bianchi, M.; Leandersson, M.; Adell, J.; Hofmann, P.; Balasubramanian, T. Kinks in the σ Band of Graphene Induced by Electron-Phonon Coupling. *Phys. Rev. Lett.* **2013**, *111*, 216806.

(26) Mazzola, F.; Polley, C. M.; Miwa, J. A.; Simmons, M. Y.; Wells, J. W. Disentangling phonon and impurity interactions in ÎŽ-doped Si(001). *Applied Physics Letters* **2014**, *104*, 173108.

(27) Mazzola, F.; Frederiksen, T.; Balasubramanian, T.; Hofmann, P.; Hellsing, B.; Wells, J. W. Strong electron-phonon coupling in the σ band of graphene. *Phys. Rev. B* **2017**, *95*, 075430.

(28) Zhang, P.; Richard, P.; Qian, T.; Xu, Y.-M.; Dai, X.; Ding, H. A precise method for visualizing dispersive features in image plots. *Review of Scientific Instruments* **2011**, *82*, 043712.

(29) Di Sante, D. et al. Flat band separation and robust spin Berry curvature in bilayer kagome metals. *Nature Physics* **2023**, *19*, 1135–1142.

(30) Mazzola, F. et al. Observation of Termination-Dependent Topological Connectivity in a Magnetic Weyl Kagome Lattice. *Nano Letters* **2023**, *23*, 8035–8042.

(31) Mazzola, F. et al. Signatures of a surface spin–orbital chiral metal. *Nature* **2024**, *626*, 752–758.

(32) Vinai, G. et al. An integrated ultra-high vacuum apparatus for growth and in situ characterization of complex materials. *Review of Scientific Instruments* **2020**, *91*, 085109.

(33) Pierantozzi, G. M. et al. Evidence of Robust Half-Metallicity in Strained Manganite Films. *The Journal of Physical Chemistry C* **2021**, *125*, 14430–14437.

(34) Panaccione, G. et al. Advanced photoelectric effect experiment beamline at Elettra: A surface science laboratory coupled with Synchrotron Radiation. *Review of Scientific Instruments* **2009**, *80*, 043105.

(35) Enkovaara, J. et al. Electronic structure calculations with GPAW: a real-space implementation of the projector augmented-wave method. *Journal of Physics: Condensed Matter* **2010**, *22*, 253202.

(36) Olsen, T. Designing in-plane heterostructures of quantum spin Hall insulators from first principles: $1\text{T}' - \text{MoS}_2$ with adsorbates. *Phys. Rev. B* **2016**, *94*, 235106.

# Uncovering the lowest thickness limit for room-temperature ferromagnetism of $Cr_{1.6}Te_2$

Sandeep Kumar Chaluvadi,[*,†] Shyni Punathum Chalil,[†,‡] Anupam Jana,[†,‡] Deepak Dagur,[†,¶] Giovanni Vinai,[†] Federico Motti,[†] Jun Fujii,[†] Moussa Mezhoud,[§] Ulrike Lüders,[§] Vincent Polewczyk,[†,∥] Ivana Vobornik,[†] Giorgio Rossi,[†,⊥] Chiara Bigi,[#] Younghun Hwang,[*,@] Thomas Olsen,[*,△] Pasquale Orgiani,[*,†] and Federico Mazzola[*,†,▽]

[†] CNR-IOM Istituto Officina dei Materiali, Area Science Park, I-34149 Trieste, Italy

[‡] International Centre for Theoretical Physics (ICTP), Str. Costiera 11, I-34151 Trieste, Italy

[¶] Department of Physics, University of Trieste, Trieste, Via Alfonso Valerio 2, 34127, Italy

[§] Normandie Univ ENSICAEN UNICAEN CNRS CRISMAT 14000 Caen, France

[∥] Groupe d'Etude de la Mati`ere Condens´ee (GEMaC), Universit´e Paris-Saclay, UMR 8635 Universit´e deVersailles Saint-Quentin en Yvelines & CNRS, Versailles, France

[⊥] Dipartimento di Fisica, Universit`a degli studi di Milano, IT-20133 Milano, Italy

[#] Synchrotron SOLEIL, F-91190 Saint-Aubin, France

[@] Electricity and Electronics and Semiconductor Applications, Ulsan College, Ulsan 44610, Republic of Korea

[△] CAMD, Computational Atomic-Scale Materials Design, Department of Physics, Technical University of Denmark, 2800, Kongens Lyngby, Denmark

[▽] Department of Molecular Sciences and Nanosystems, Ca' Foscari University of Venice, I-30172 Venice, Italy

E-mail: chaluvadi@iom.cnr.it ; younghh@uc.ac.kr ; tolsen@fysik.dtu.dk ; pasquale.orgiani@cnr.it ; federico.mazzola@unive.it

**Phase Identification:**

**Table S1:** Lattice constants for various $Cr_xTe_y$ compositions

| $Cr_xTe_y$ Compounds | a (Å) | b (Å) | c (Å) | Reference |
|---|---|---|---|---|
| CrTe | 3.997 | 3.997 | 6.222 | [1][2][3] |
| $CrTe_2$ | 3.93 | 3.93 | 6.02 | [4] |
| $Cr_{1.16}Te_2$ ($Cr_{9.35}Te_{16}$) | 7.8508 | 7.8508 | 11.8372 | [5] |
| $Cr_{1.24}Te_2$ ($Cr_{0.62}Te$) | 7.792 |  | 11.980 | [6] |
| $Cr_{1.25}Te_2$ ($Cr_5Te_8$) | 13.5201 | 7.8205 | 11.9791 | [7][8] |
| $Cr_{1.33}Te_2$ ($Cr_2Te_3$) | 6.814 | 6.814 | 12.073 | [9][10] |
| $Cr_{1.6}Te_2$ ($Cr_4Te_5$) | 6.557 | 6.557 | 12.495 | [11] |
| | 6.87±0.2 | 6.87±0.2 | 12.496 ± 0.003° | This work |
| $Cr_{1.66}Te_2$ ($Cr_5Te_6$) | 6.913 | 3.970 | 12.44 | [9] |

Here, we present the table with various compositions of $Cr_xTe_y$ alongside their corresponding lattice constants. Notably, the data reveals a trend where increasing Cr-intercalation in the parent compound ($CrTe_2$) leads to expansion in the out-of-plane lattice constant, denoted as 'c.' Importantly, our measured lattice constants derived from STM and XRD analysis (a ≈ 6.87±0.2 Å, c ≈ 12.495 Å) for the composition $Cr_{1.6}Te_2$ ($Cr_4Te_5$) align with literature values, affirming the phase of our system.

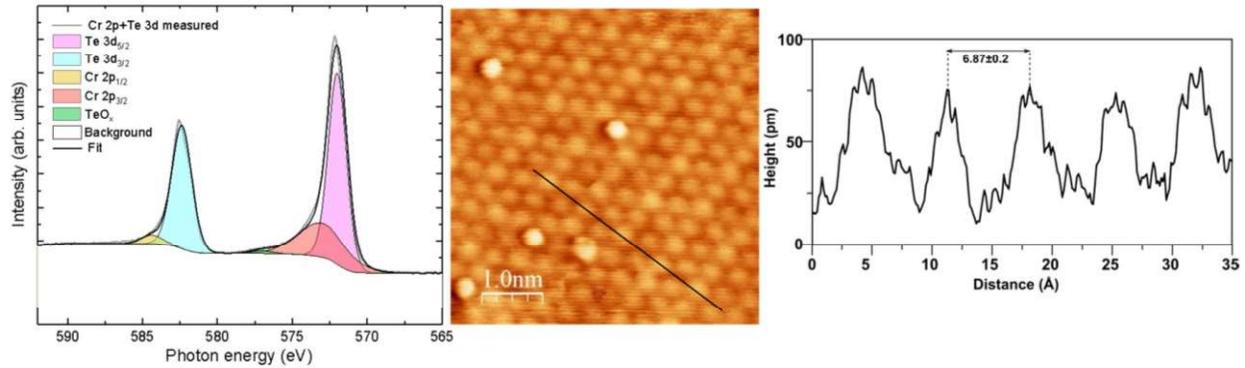

**Figure S1:** (left) XPS spectra of Cr2p + Te 3d core-levels taken at the photon excitation energy of 800 eV, (middle, right) STM atomic-resolution and corresponding line profile with an in-plane lattice parameter measured ~6.87±0.2 Å.

Furthermore, XPS measurements performed at the APE-HE beamline at Elettra synchrotron are shown in Fig. S1. The XPS core-level spectra of Cr 2p and Te 3d of the in-situ transferred 75 nm thick $Cr_{1.6}Te_2$ film, taken at the photon excitation energy of 800 eV. In order to estimate the Cr/Te ratio in the film, Cr 2p + Te 3d core-level spectra were fitted using mixed Gaussian-Lorentzian line shapes after subtracting the background intensity. The background was due to the presence of secondary electrons using the Shirley background and considering their cross-section profiles. The composition of film obtained was close to ~$Cr_{1.55}Te_2$.

**Rocking Curve:**

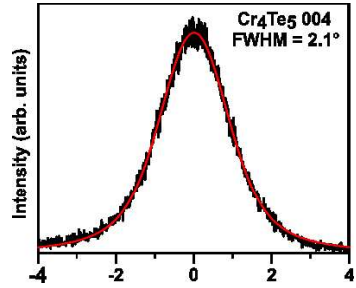

**Figure S2:** The full-width at half maximum (FWHM) deduced from the rocking curve analysis for the 004 reflection is approximately 2.1°. Comparing this value to those reported in the literature for similar phase compounds, we find that our films exhibit approximately 50% lower broadening. The literature values for FWHM range from about 3.21° [12] to 3.3° [11], indicating broader peaks compared to our findings. This discrepancy in broadening can be attributed to various factors, including the van der Waals nature of the material and the weak interaction of ionic bonds.

**Thin film growth optimization:**

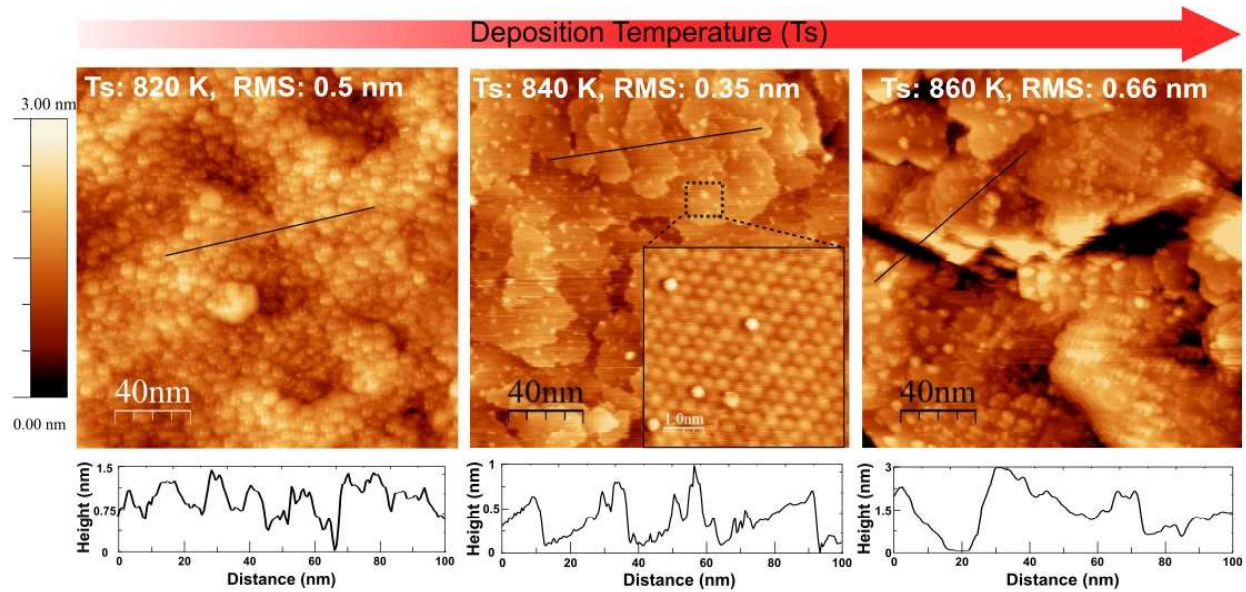

**Figure S3:** STM topography images of $Cr_{1.6}Te_2$ thin films grown at different temperatures 820, 840 and 860 K in the scan area of 200×200 nm$^2$ and their corresponding line profiles respectively. The atomic-resolution of the film deposited at 840 K is also shown in the inset.

The c-plane oriented single-crystalline $Al_2O_3$ substrates were commercially purchased from the SurfaceNet Gmbh, Germany. The as-received substrates were subsequently diced into 5×5 mm$^2$ dimensions, rinsed in ethanol and dried under $N_2$ flow, introduced into the PLD uhv chamber with the base pressure of 5×10$^{-8}$ mbar. $Cr_{1.6}Te_2$ thin films were deposited in uhv onto $Al_2O_3$ substrates by using 1$^{st}$ harmonic Nd:YAG laser source with a repetition rate of 2 Hz, respectively. The crystalline structure of the deposited films was assessed via X-ray diffraction (XRD), following which, the substrate deposition temperature ($T_s$) was finely tuned for obtaining flat, defect free surface. This optimization process was facilitated by in-situ scanning tunneling microscopy (STM) surface analysis, allowing for precise control over the growth conditions.

Upon evaluating the surface characteristics of the deposited films, it was observed that films deposited at 820 K exhibited a randomly arranged atomic species, accompanied by a root-mean-square (RMS) roughness value of 0.5 nm. The corresponding line profile shows a tiny cluster of islands dispersed throughout the surface. Similar surface morphology was also reported in [11], [12] however, their surface roughness was three times higher than the one observed in our case. Furthermore, the surface morphology of films grown at 840 K displayed clear step-like features, effectively reducing RMS roughness (0.35 nm) and absence of droplets. Detailed line profile revealed flat surfaces with terrace widths and step heights ranging between 25-30 nm and 0.5-0.6 nm, respectively. Inset shows the atomic-resolution with defect free surface. A further increase in Ts to 860 K resulted in increased surface roughness and the formation of deep holes and clustering of atoms on the surface.

Our findings highlight the delicate balance between substrate deposition temperature and surface morphology in epitaxial growth processes. Through meticulous optimization, we were able to achieve significant improvements in surface flatness and morphology, reminiscent of previous observations in other thin film systems [13]. This underscores the critical role of precise temperature control in dictating the surface characteristics of transition metal dichalcogenides (TMDs) thin films.

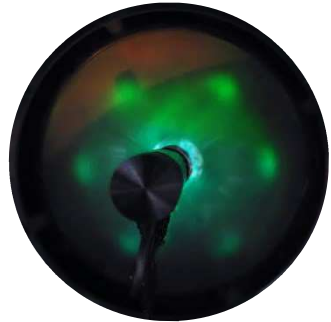

**Figure S4:** Low-energy electron diffraction (LEED) pattern of the $Cr_{1.6}Te_2$ film taken at 60 eV. It displays hexagonal symmetry thus confirming the observed 2D FFT image as shown in Fig 1. in the main text.

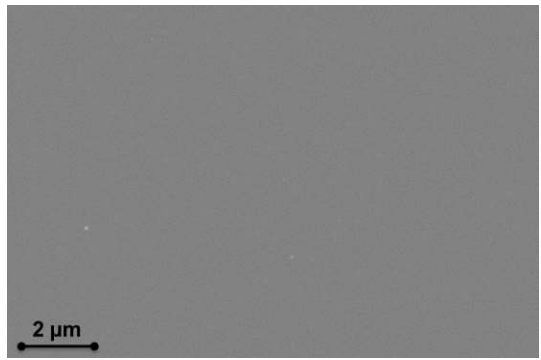

**Figure S5:** SEM micrograph of the $Cr_{1.6}Te_2$ thin film recorded at a magnification of 20kx. The microstructural surface quality of the $Cr_{1.6}Te_2$ thin film deposited by first harmonic Nd:YAG laser is assessed by scanning electron microscopy. The film grows uniformly over large areas without any formation of voids or droplets on the film surface. Therefore, it can be scaled easily for large area device applications.

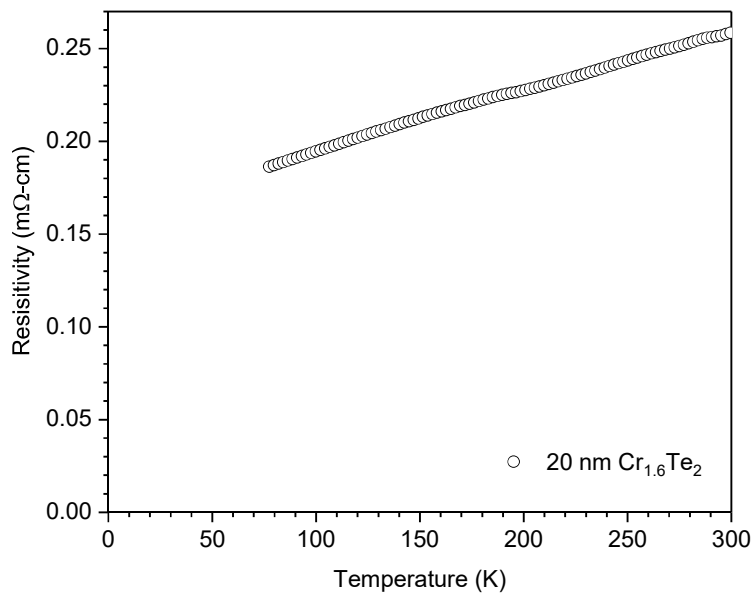

**Figure S6:** Temperature-dependent electrical transport measurement of the 20 nm $Cr_{1.6}Te_2$ film. It shows metallic nature with a room-temperature resistivity of ~0.25 mΩ-cm.

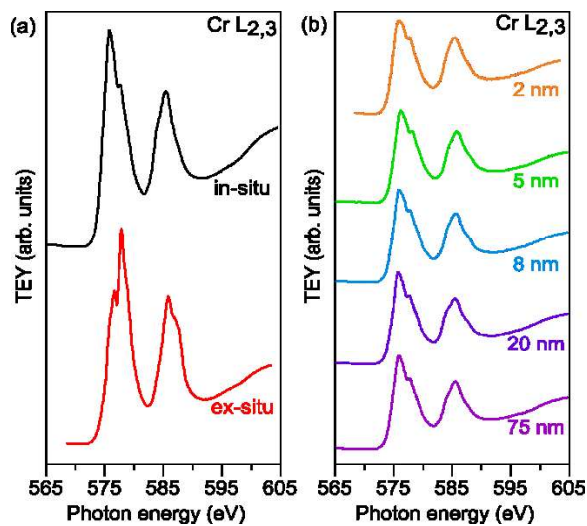

**Figure S7:** (a) Comparison of the Cr $L_{2,3}$ absorption edge of an in-situ (black curve) and ex-situ (red curve) sample of $Cr_{1.6}Te_2$ film. (b) Thickness-dependent in-situ XAS from the bulk-like 75nm down to ultra-thin films of 2 nm.

To determine the differences in the Cr $L_{2,3}$ absorption edge, XAS experiments were performed on both *in-situ* as well as on *ex-situ* transferred samples at room-temperature (300 K) (Figure S7). The *in-situ* sample was transferred under UHV conditions (<$10^{-10}$ mbar) to the APE-HE end station and the measurements are acquired in total electron yield (TEY) mode. There are clear spectral feature differences between the in-situ and ex-situ transferred samples. The observed ex-situ Cr $L_{2,3}$ XAS spectrum is comparable to that of $Cr_2O_3$, showing that the surface of the $Cr_{1.6}Te_2$ film has been oxidized.

The shape of the Cr L2,3 XAS edges looks similar from bulk-like 75 nm down to 2nm films indicating the electronic properties of the films are stable with thickness. At higher photon energies (> 600 eV), the observed broad hump belongs to the Te M edges. As both Cr 2p and Te 3d absorption edges fall close to each other, an overlap in the spectra can be expected. However, due to the negligible cross-section of Te M edges (< 5%) with respect to the L edges of Cr, very small contribution may arise from the Te M edge, which can be ignored [10].

**Magnetic characterization:**

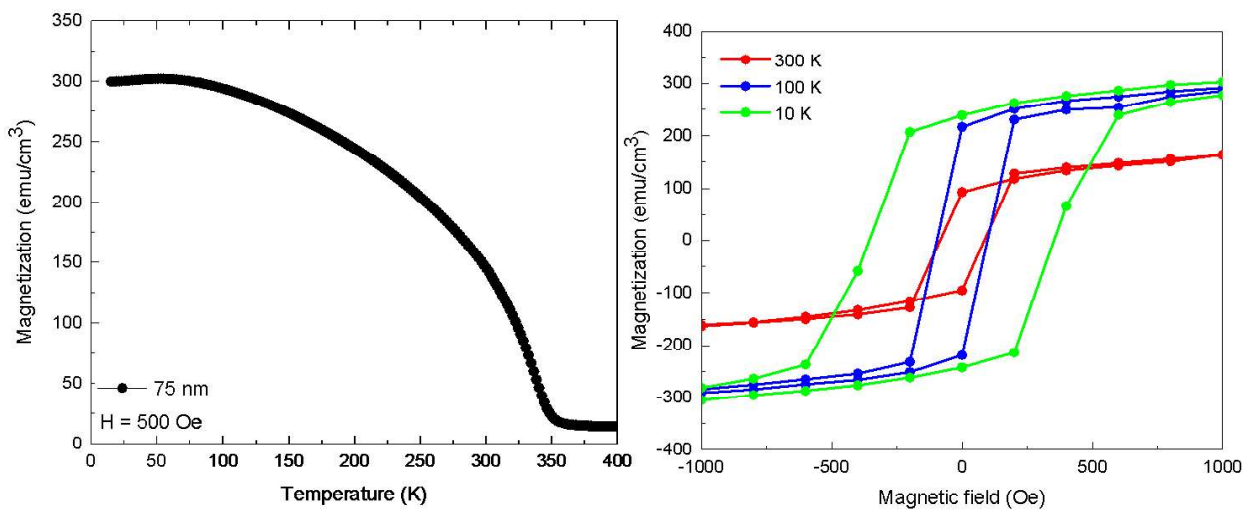

**Figure S8:** (left) Temperature-dependent magnetization curve of the 75 nm thick $Cr_{1.6}Te_2$ film obtained with SQUID clearly shows that the Curie temperature (Tc) lies above 350 K; (right) Hysteresis loops taken at the different temperatures. Note that the remanence magnetization (Mrem) values at 100 and 300 K are almost doubled, which is in correspondence with the XMCD% of the film as described in the main text.

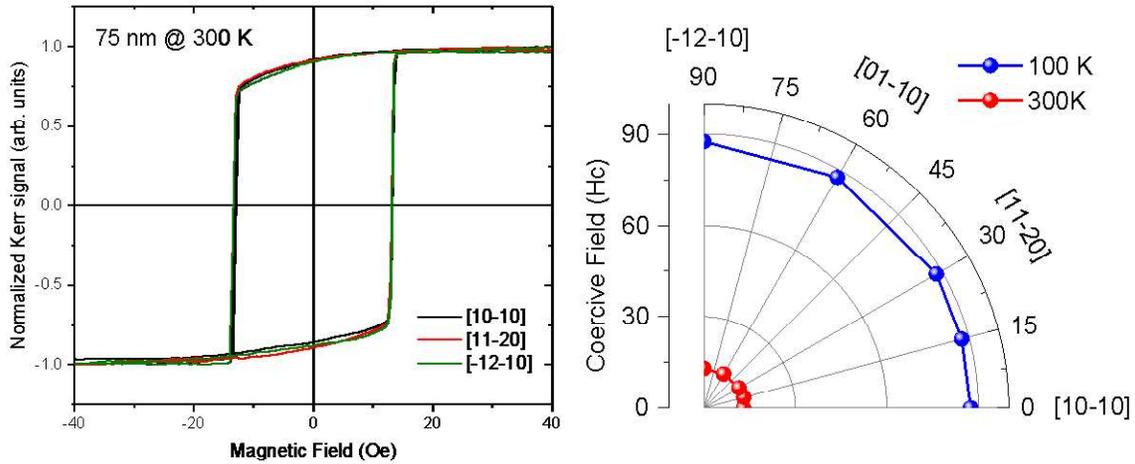

**Figure S9:** (left) *In-plane* Magneto-optical Kerr hysteresis loops taken along different crystallographic directions of the 75 nm thick $Cr_{1.6}Te_2$ film at 300 K. All the hysteresis loops have similar coercive and remanence fields i.e., the film is magnetically isotropic within the plane. (right) Angular dependent coercive fields of the same film taken at 100 and 300 K, which confirms the isotropic nature of the film.

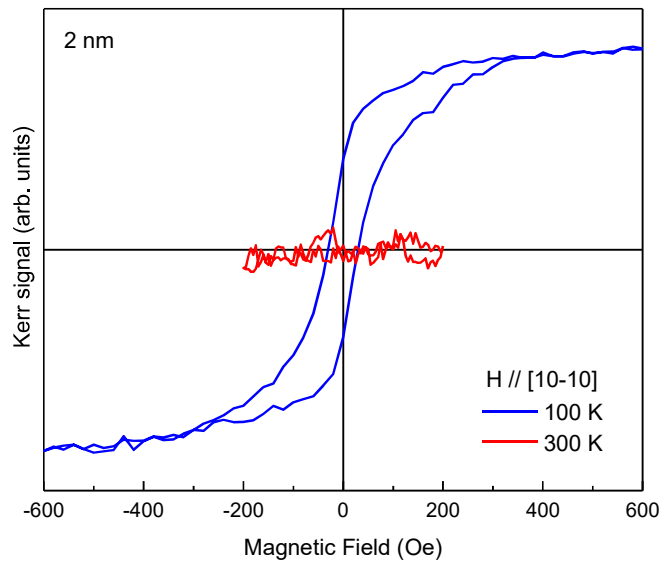

**Figure S10:** In-plane MOKE hysteresis loops for the 2nm $Cr_{1.6}Te_2$ film measured along the [10-10] crystallographic direction at two temperatures (300 and 100 K). At 300 K, the MOKE signal remains nearly flat, indicating absence of ferromagnetic ordering, whereas at 100 K, a distinct hysteresis loop emerges, indicative of ferromagnetic ordering.

**Details on DFT calculation:**

The structure used for the DFT calculations is shown Fig. S11. It is comprised of van der Waals bonded bulk $CrTe_2$ with three intercalated Cr atoms for every five formula units of $CrTe_2$. The structure was thus constructed using a $\sqrt{5} \times \sqrt{5}$ super cell. All DFT calculations were performed with the electronic structure package GPAW [14] using the projector-augmented wave method and plane waves. For the relaxation of the $Cr_{1.6}Te_2$ structure we used a plane wave cutoff of 800 eV and a 7x7 uniform k-point sampling. The structure was found to reside 0.03 eV above the convex hull and is thus highly stable. The calculations yielded a metallic ground state with 13.9 $\mu_B$ per formula unit, with the local moments of the intercalated Cr atoms being slightly larger than the rest.

The band structure was unfolded to the Brillouin zone of the pristine $CrTe_2$ structure using the method of Ref. [15] with spin-orbit coupling (SOC) included. Due to the heavy Te atoms SOC yields a strong modification of the band structure as shown in Fig. S12. In particular, the valence band maximum is raised by roughly 0.4 eV such that it straddles the Fermi level when SOC is included. Comparing with Fig. 3 of the main text it is clear that inclusion of SOC is crucial in order to obtain agreement with the ARPES measurements.

For the calculations a c-axis parameter of 6.25 Å was used. Such a value, which is half of what reported by the XRD, could explain the absence of any diffraction peak with odd Miller's index, which remain too weak to be detected (we are expecting that the real structure is half of the XRD result indeed).

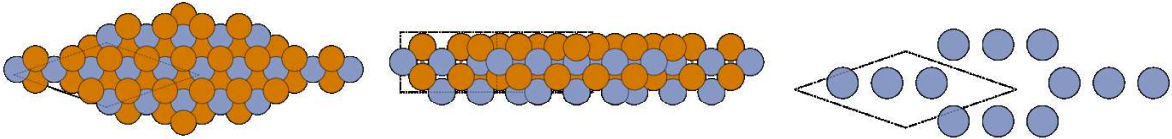

**Figure S11**: The structure used to simulate $Cr_{1.6}Te_2$. Left: top view of the $\sqrt{5} \times \sqrt{5}$ structure (in a showing a $2 \times 2$ repetition of the cell). Middle: side view of the structure. Right: top view of the intercalated atoms.

Finally, the full projected density of states (within the experimental range considered it has been presented in the main text) is also represented in Fig. S13, where the spin majority and minority characters have been both shown. In addition, in Fig. S13, we show the orbital projected nature of the electronic structure where one can see the prominent Cr character of the electronic structure studied within the range captured by our experiment.

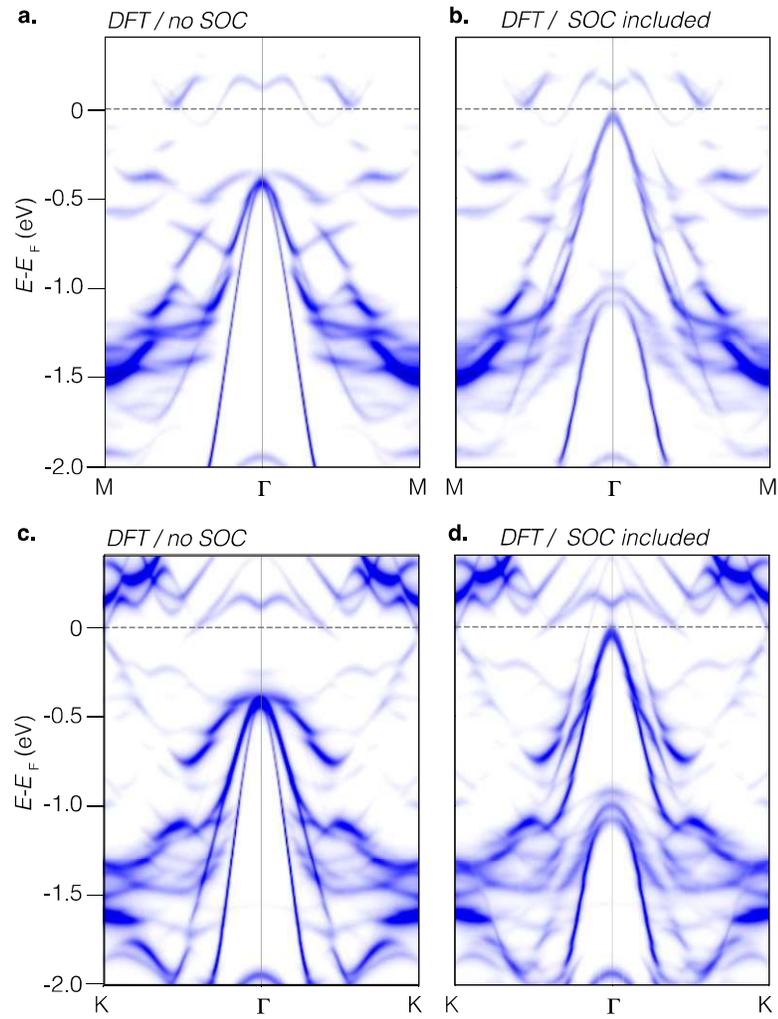

**Figure S12:** Effect of SOC. **a.** Electronic structure along the M-Γ-M direction without and **b.** with SOC. **c.** Electronic structure along the K-Γ-K direction without and **d.** with SOC. In both directions, the effect of SOC is a drastic change of the electronic structure, in terms of energy shift, renormalization, spectral weight distribution, and general shape.

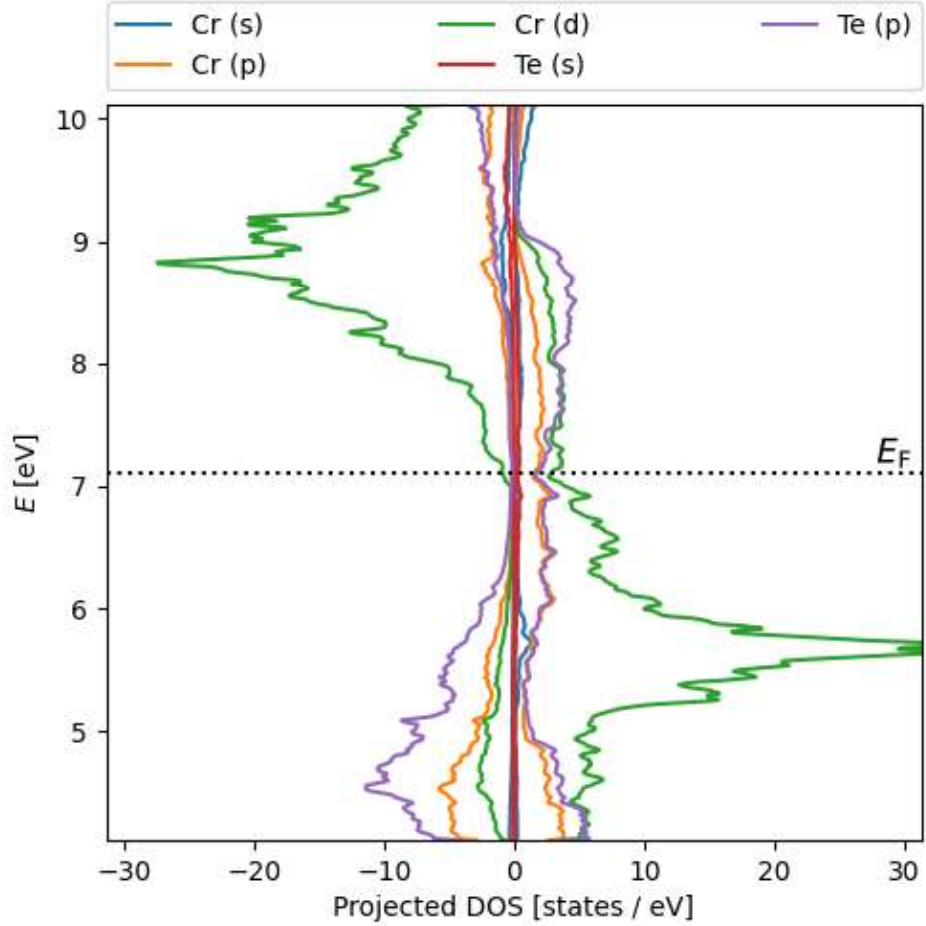

**Figure S13:** Spin Projected density of sates resolved for each orbital character. As one can see, the d orbitals of Cr are the main responsible for the ferromagnetic character of the electronic structure. A significant contribution to the spin character is also visible in the p orbitals of Te but at energies much lower (>2 eV) below the Fermi level.

**Sum rules analysis:**

An element-sensitive quantitative evaluation on the magnetic spin and orbital moments has been obtained on the 75 nm thick $Cr_{1.6}Te_2$ sample via sum rules analysis [16] taking into account the fact that, being Cr an early transition metal, the evaluation of the spin moment contribution is delicate due to the partial overlap between $L_2$ and $L_3$ edges [16], [17]. The calculated integrals and obtained sum rules parameters are shown in Figure SX for the 100 K case. Spin and orbital moments were then calculated by using the formulas

$$m_{spin} = \frac{-(6p-4q)(10-n_{3d})}{r} C \quad m_{orb} = \frac{-4q(10-n_{3d})}{3r}$$

with $n_{3d}$ equal to 2.5 for $Cr_{1.6}Te_2$ and $C$ equal to 2 as correction factor for Cr spin moment [18].
The obtained total magnetic moment ($m_{tot} = m_{spin} + m_{orb}$) is of $\approx (2.02 \pm 0.5)$ $\mu_B$ per Cr atom at 100 K, and $\approx (1.05 \pm 0.2)$ $\mu_B$ per Cr atom at room temperature. The value obtained at 100 K is consistent with what observed in thick $Cr_3Te_4$ intercalated films at low temperature [19], and its lower value compared to the theoretically expected 3.2 $\mu_B$/Cr atom can be attributed to the not yet fully reached magnetic saturation at that temperature, and to the intrinsic ambiguity of the determination of $m_{spin}$. On the other hand. the relatively large moment persisting at 300 K confirms the solid ferromagnetic behavior at room temperature. For both temperatures, the obtained values of $m_{orb}$ are quite small (0.05 and 0.017 $\mu_B$ at 100 and 300 K respectively), suggesting a quenched Cr orbital moment as observed in other $Cr_{1-\delta}Te$ compounds [18]–[20].

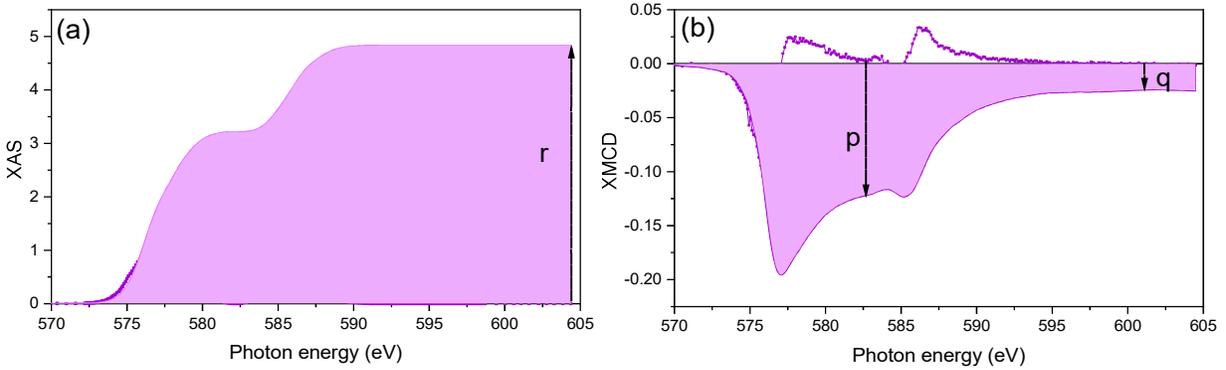

**Figure S14:** (a) XAS and (b) XMCD spectra at Cr $L_{2,3}$ edges for 75 nm $Cr_{1.6}Te_2$ thin film at 100 K. A step-like function was employed to exclude the background non-magnetic contribution.

### 75 nm film magnetism

The 75 nm thick film shows an XMCD value of 11% at 100 K, which is almost twice the room temperature value (6%). This result is in agreement with the temperature-dependent magnetic saturation measurements performed by SQUID (shown above), in which the saturation magnetization is doubled from 100 K to 300 K. Additionally, SQUID temperature dependencies show a Curie temperature close to 350 K for 75 nm thick films.

**Electronic dimensionality of the system**

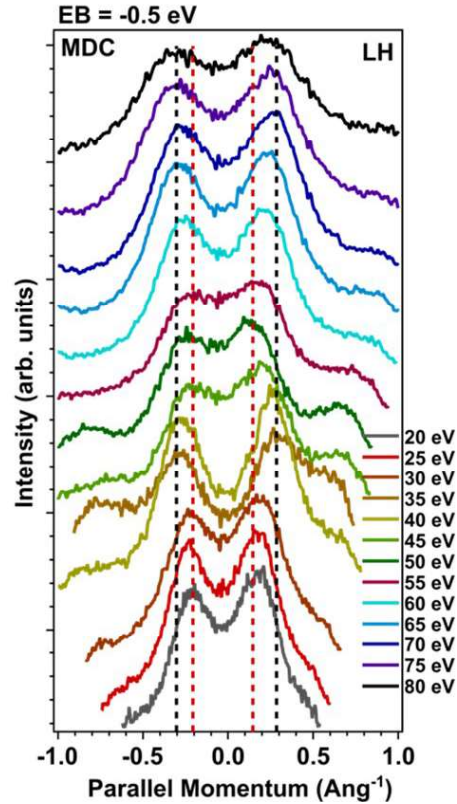

**Figure S15:** Here, we show momentum distribution curves extracted at the Fermi level position for several values of photon energy used. As one can see, apart from a general redistribution of the spectral weight, the centroid of the peaks does not change significantly. This behavous is consistent with a two-dimensional chacracter of the relevant bands.

**References:**
[1] Y. Liu, S. K. Bose, and J. Kudrnovský, "First-principles theoretical studies of half-metallic ferromagnetism in CrTe," *Phys. Rev. B - Condens. Matter Mater. Phys.*, vol. 82, no. 9, pp. 1–8, 2010.
[2] M. Wang *et al.*, "Two-dimensional ferromagnetism in CrTe flakes down to atomically thin layers," *Nanoscale*, vol. 12, no. 31, pp. 16427–16432, 2020.
[3] S. Sun *et al.*, "Anisotropic magnetoresistance in room temperature ferromagnetic single crystal CrTe flake," *J. Alloys Compd.*, vol. 890, p. 161818, 2022.
[4] Y. Ou *et al.*, "ZrTe2/CrTe2: an epitaxial van der Waals platform for spintronics," *Nat. Commun.*, vol. 13, no. 1, pp. 1–9, 2022.
[5] G. Cao *et al.*, "Structure, chromium vacancies, and magnetism in a C r12-x T e16 compound," *Phys. Rev. Mater.*, vol. 3, no. 12, pp. 1–8, 2019.
[6] Y. Liu and C. Petrovic, "Critical behavior of the quasi-two-dimensional weak itinerant ferromagnet trigonal chromium telluride Cr0.62Te," *Phys. Rev. B*, vol. 96, no. 13, pp. 1–7, 2017.
[7] X. Zhang, T. Yu, Q. Xue, M. Lei, and R. Jiao, "Critical behavior and magnetocaloric effect in monoclinic Cr5Te8," *J. Alloys Compd.*, vol. 750, pp. 798–803, 2018.

[8]  Z.-L. Huang, W. Kockelmann, M. Telling, and W. Bensch, "A neutron diffraction study of structural and magnetic properties of monoclinic Cr 5 Te 8," 2007.

[9]  A. F. Andresen, E. Zeppezauer, T. Boive, B. Nordström, and C.-I. Brändén, "The Magnetic Structure of Cr2Te3, Cr3Te4, and Cr5Te6.," *Acta Chemica Scandinavica*, vol. 24. pp. 3495–3509, 1970.

[10]  D. M. Burn *et al.*, "Cr2Te3 Thin Films for Integration in Magnetic Topological Insulator Heterostructures," *Sci. Rep.*, vol. 9, no. 1, pp. 1–10, 2019.

[11]  J. Wang *et al.*, "Epitaxial growth and room-temperature ferromagnetism of quasi-2D layered Cr4Te5thin film," *J. Phys. D. Appl. Phys.*, vol. 55, no. 16, 2022.

[12]  W. Wang *et al.*, "Fabrication and magnetic-electronic properties of van der Waals Cr4Te5ferromagnetic films," *CrystEngComm*, vol. 24, no. 3, pp. 674–680, 2022.

[13]  S. K. Chaluvadi *et al.*, "Pulsed laser deposition of oxide and metallic thin films by means of Nd:YAG laser source operating at its 1st harmonics: recent approaches and advances," *J. Phys. Mater.*, vol. 4, no. 3, p. 032001, 2021.

[14]  J. Enkovaara *et al.*, "Electronic structure calculations with GPAW: a real-space implementation of the projector augmented-wave method," *J. Phys. Condens. Matter*, vol. 22, no. 25, p. 253202, Jun. 2010.

[15]  V. Popescu and A. Zunger, "Extracting E versus k effective band structure from supercell calculations on alloys and impurities," *Phys. Rev. B*, vol. 85, no. 8, p. 085201, Feb. 2012.

[16]  C. T. Chen *et al.*, "Experimental confirmation of the x-ray magnetic circular dichroism sum rules for iron and cobalt," *Phys. Rev. Lett.*, vol. 75, no. 1, pp. 152–155, 1995.

[17]  W. L. O'Brien, B. P. Tonner, G. R. Harp, and S. S. P. Parkin, "Experimental investigation of dichroism sum rules for V, Cr, Mn, Fe, Co, and Ni: Influence of diffuse magnetism," *J. Appl. Phys.*, vol. 76, no. 10, pp. 6462–6464, Nov. 1994.

[18]  A. I. Figueroa *et al.*, "Magnetic Cr doping of Bi2Se3: Evidence for divalent Cr from x-ray spectroscopy," *Phys. Rev. B*, vol. 90, no. 13, p. 134402, Oct. 2014.

[19]  R. Chua *et al.*, "Room Temperature Ferromagnetism of Monolayer Chromium Telluride with Perpendicular Magnetic Anisotropy," *Adv. Mater.*, vol. 33, no. 42, p. 2103360, Oct. 2021.

[20]  X. Zhang *et al.*, "Room-temperature intrinsic ferromagnetism in epitaxial CrTe2 ultrathin films," *Nat. Commun. 2021 121*, vol. 12, no. 1, pp. 1–9, May 2021.